\documentclass[eqsecnum,preprintnumbers,amsmath,amssymb]{revtex4}
\usepackage{graphicx}% Include figure files
\usepackage{dcolumn}% Align table columns on decimal point
\usepackage{bm}% bold math
\begin{document}
\title{Mutual friction in helium II: a microscopic approach}
\author{H. M. Cataldo}                     
\email{cataldo@df.uba.ar}
\homepage{http://www.df.uba.ar/users/cataldo}
\affiliation{Consejo Nacional de Investigaciones Cient\'{\i}ficas y T\'ecnicas\\and
Departamento de F\'{\i}sica, Facultad de Ciencias Exactas y Naturales, \\
Universidad de Buenos Aires, RA-1428 Buenos Aires, Argentina}

\begin{abstract}
We develop a microscopic model of mutual friction
represented by the
dissipative dynamics of a normal fluid flow which interacts with the helical normal modes of
vortices comprising a lattice in thermal equilibrium.
Such vortices are assumed to
interact with the quasiparticles forming the
normal fluid through a pseudomomentum-conserving scattering Hamiltonian. 
We study the approach to equilibrium of the normal fluid flow for temperatures below 1 K,
deriving an equation of motion for the quasiparticle pseudomomentum which leads to
the expected form predicted by the HVBK equations.
We obtain an expression for 
the mutual friction coefficient $B$ in terms of
microscopic parameters, which
turns out to be practically independent of
the vortex mass for values arising from diverse theories. 
By comparing
 our expression of $B$ with previous theoretical estimates,
we deduce interesting qualitative
features about the excitation of Kelvin modes by the quasiparticle scattering.

\end{abstract}
\maketitle
\section{Introduction}\label{s1}
%Below the lambda transition, at temperatures below 
%$T_\lambda=2.172$ K, liquid helium is known as {\it helium II} and
%can be regarded as a mixture
%of a normal fluid with mass density $\rho_n$ and a superfluid with mass density $\rho_s$.
%Such densities are temperature dependent, so that $\rho_s(T_\lambda)=\rho_n(0)=0$,
%while the total density $\rho=\rho_s+\rho_n$ remains nearly constant. The most striking 
%property of such a superfluid component is, perhaps, that it can only rotate through vortices
%having microscopic cores and quantized circulations. In fact, the
%circulation of the superfluid velocity field around each of such vortices is quantized in
%units of $h/m_4$, the so-called quantum of circulation $\kappa=h/m_4$, given by the ratio of 
%Planck's constant and the mass of one $^4$He atom. 
When a sufficiently fast rotating sample of liquid helium is cooled below the lambda
temperature, all the rotation of the superfluid becomes concentrated in 
a uniform array of quantized vortex filaments 
parallel to the axis of rotation \cite{don,bar}.
By contrast,
the macroscopic superfluid velocity field, corresponding to spatial averages
over regions large compared with the spacing between vortices,
yields the usual configuration of solid body flow,
${\bf v}_s({\bf r})=\Omega_{\rm rot}{\bf \hat z}\times{\bf r}$
for a rotation frequency $\Omega_{\rm rot}$ around the $z$ axis.
Just as the superfluid flow is microscopically formed by vortices, the normal fluid consists of
superfluid quasiparticle excitations, phonons and rotons, the average flow of which is 
characterized
by the normal fluid velocity field ${\bf v}_n$. 
In equilibrium both fluids move with the same velocity
${\bf v}_n={\bf v}_s$ and such a behavior arises, from a microscopic viewpoint, from the vortex
motion with the normal fluid velocity in order to avoid dissipation.
That is, any relative motion of vortices with
respect to the normal fluid in their vicinity, is subjected to a friction force that
causes such a motion to eventually cease. Such a {\it mutual friction force} \cite{hall} between the two 
fluids plays then a central role
in the mechanism which maintains the stability of the above equilibrium state.
A well-known phenomenological model for this macroscopic dynamics is represented by the so-called
Hall-Vinen-Bekharevich-Khalatnikov (HVBK) equations \cite{hall,bek},
basically consisting of a Navier-Stokes equation for the normal fluid and an Euler equation
for the superfluid, which, in the absence of pressure and temperature gradients, are coupled together
only by mutual friction.
There is a simple configuration which allows to show the basic features of this process, namely
rectilinear flows of uniform vorticity \cite{jpa1}, 
\begin{eqnarray}
{\bf v}_s({\bf r},t)&=&-2\,\Omega_s(t)\,y\,{\bf \hat x}\label{1.1}\\
{\bf v}_n({\bf r},t)&=&-2\,\Omega_n(t)\,y\,{\bf \hat x}\label{1.2}
\end{eqnarray}
($y<0$), where $\Omega_s(t)$ and $\Omega_n(t)$ should converge for $t\rightarrow\infty$ to 
a steady state value. Then, to make contact with the standard rotational configuration,
we may identify such a value with the former angular velocity $\Omega_{\rm rot}$.
Note that this assignment leads to a uniform vorticity ${\bf \nabla}\times{\bf v}_s
=2\,\Omega_{\rm rot}\,{\bf \hat z}$ which coincides with that of the rotational scheme.
The HVBK equations for such flows are very simple and read
\begin{equation}
\rho_n\frac{\partial{\bf v}_n}{\partial t}  =  {\bf F}
=-\rho_s\frac{\partial{\bf v}_s}{\partial t},
\label{p1}
\end{equation}
where $\rho_n$ and $\rho_s$ denote the normal fluid and superfluid mass densities, respectively, and
the mutual friction force ${\bf F}$ can be written for temperatures below 1 K as \cite{jpa1},
\begin{equation}
{\bf F}=-B\rho_n\,\Omega_{\rm rot}
({\bf v}_n-{\bf v}_s),
\label{p3}
\end{equation}
being  $B$ a
dimensionless dissipative coefficient.
Such a low temperature regime corresponds to $\rho_n\ll\rho_s$, 
which, according to (\ref{p1}), implies that the main
time dependence should lie within the normal fluid velocity. This suggests that a suitable
approach to the problem may consist in regarding the superfluid component as a thermal
equilibrium heat bath which interacts with a nonequilibrium normal fluid flow.
Keeping such a picture as our basic premise,
we shall analyze in the present paper a microscopic model of mutual friction, which reproduces
the main features of the above macroscopic dynamics, yielding an explicit expression of
$B$ as a function
of microscopic parameters.

The microscopic basis of mutual friction remains as one of the most intrincate problems of 
superfluidity. In such a context,
theoretical approaches may strongly differ, even in 
significant questions such as the
 existence of a nondissipative component of the mutual friction
force \cite{sonao}, which is absent from our modelling \cite{jpa1}.
A better understanding of the microscopic principles governing mutual friction 
would also contribute to
clarify important issues on the subject of quantum turbulence at finite temperature \cite{turb}. 
In fact, it is just the mutual friction force which accounts for the strong locking between superfluid and
normal fluid along the turbulent cascade, where recent simulations have shown that
the residual slip velocity ${\bf v}_s-{\bf v}_n$ plays a central role \cite{roche}. 
In addition, such simulations suggest that the cross-over between 
zero-temperature and finite temperature quantum turbulence occurs at a 
lower temperature than the usual estimation of 1 K, 
hence partially placing the latter regime within 
the temperature range of the present investigation.

Another important source of controversy arises from the mass of quantized vortices.
On the one hand, many works have considered it as a negligible
parameter under the 
assumption that it should be equivalent to the hydrodynamic mass of a
core of atomic dimensions \cite{don}.
Another theories, however, 
yield several orders of magnitude higher values for the vortex mass, 
casting doubt on models based on massless vortices  \cite{duan}.
Moreover, it has been argued that an unambiguous vortex mass may not exist,
and that inertial effects in vortex dynamics may be scenario-dependent \cite{thou}.
Finally, we should also mention that
 there have been conflicting results for the vortex mass in superconductors as well \cite{ao}.
A possible way out to such uncertainties has been recently suggested 
based on the concept of {\em pseudomomentum}.
In fact, just as the momentum corresponds to the generator of ordinary translations, the so-called
pseudomomentum  generates translations which keep an eventual background medium fixed \cite{peie}.
Thus, the pseudomomentum often appears as a useful tool in fluid mechanics \cite{b-m}.
In the present case we shall concentrate our attention on the pseudomomentum of quasiparticles forming
tbe normal fluid and on the pseudomomentum of vortices, being the background medium the superfluid.
 The quasiparticle pseudomomentum per unit volume turns out to be simply
$\rho_n ({\bf v}_n-{\bf v}_s)$, which may be easily interpreted as the ordinary momentum
of a normal fluid moving relatively to the superfluid. On the other hand, 
the vortex
pseudomomentum requires a more careful treatment, recently 
shown to lead to an alternative approach to the dissipative 
vortex dynamics, free of the ambiguities carried by the uncertainty of the vortex mass \cite{jpa1}.
Our present study of the dissipative normal fluid dynamics
will utilize the concept of pseudomomentum as an important tool,
finding again results which are
practically independent of the vortex
mass for a wide range of values.

The dynamics of the normal fluid flow is far less known than that of the superfluid, because,
apart from very recent efforts \cite{paol},
there is a lack of experimental observation techniques at these low temperatures.
Only recently, theoretical research has been focused on this issue by means of numerical 
simulations, which provided valuable information about normal fluid flow patterns due to 
the mutual friction interaction \cite{kivo}. Another interesting aspect, which has been recently 
investigated,
 concerns the stability characteristics of the normal fluid flow under
mutual friction forcing from the superflow \cite{god}.

Our approach will consist in 
assuming a heat bath formed by a vortex
lattice in thermal equilibrium, which interacts with a quasiparticle flow. The dominant contribution
to the heat capacity of such a lattice should arise from the thermal excitation of
helical waves, corresponding to effectively independent vortices \cite{fet1}.
The role of such oscillations in mutual friction has
 been scarcely treated in the literature.
We are only able to mention a couple of papers \cite{fet2,son}, 
that long ago reached the conclusion that the
damping of vortex oscillations due to phonon scattering, should not modify appreciably 
the value of the friction coefficient calculated for a rigid vortex.
The same conclusion was recently obtained for a high-frequency branch of
 helical waves, within a wider temperature range, including a roton-dominated regime \cite{jpa}.

In building a theory with massive vortices,
one can readily make use of a close analogy with the well-known electrodynamical problem of
a point charge subjected to magnetic and electric fields \cite{yosh}.
Particularly, the quantization
of the theory, which greatly simplifies the treatment when
the scattering excitation of
vortex waves is taken into account \cite{fet2},
 arises immediately from this analogy.
Such an analogy also leads to an immediate identification of the 
 vortex pseudomomentum, allowing us to
build a proper form for a pseudomomentum-conserving  scattering Hamiltonian.

This paper is organized as follows.
In the next section, we propose a Hamiltonian
model for the interaction of a vortex line
 with a quasiparticle gas and
a background superflow of uniform vorticity.
 In Sec.~\ref{s3} we obtain the equation of motion for the quasiparticle pseudomomentum,
 which leads to the expected form predicted by 
the HVBK equations. In Sec.~\ref{s4} we study our expression for
 the mutual friction coefficient $B$ comparing
this result with previous theoretical estimates. 
Finally in Sec.~\ref{s7} we give our concluding remarks.

\section{Hamiltonian model}\label{s2}
\subsection{Vortex Hamiltonian}\label{sec1A}
Let us consider an otherwise rectilinear vortex filament performing helical oscillations about its
unperturbed position parallel to the $z$ axis \cite{don}. 
The wavelength $\lambda$ is supposed to be much greater than
the amplitude (radius of the helix), and to have a full description of the helix, one should
also know the direction (right or left) of the helical deformation, or equivalently, the 
direction of the wave vector $k{\bf \hat z}$ ($k=\pm 2\pi/\lambda$).
Thus, it suffices to know the location
of the vortex core at a given plane normal to the $z$ axis, say $z=0$, to have a 
full determination of the position of the whole vortex filament.
Periodic boundary conditions over a length $L$ (vortex line length)
 along the $z$ axis 
determine the possible values of the wave vector as $k=2\pi m/L$, where $m$ is an integer.
The vortex core position ${\bf r}$  may then be written as a summation
over generalized two-dimensional
coordinates ${\bf r}_k$ associated to normal modes labeled by the wave vector $k{\bf \hat z}$.
We shall first consider true oscillatory modes ($k\neq 0$), since
modes with $k=0$ corresponding to rigid displacements of the vortex filament require
a separate treatment \cite{jpa1}.
The equation of motion for ${\bf r}_k(t)$
is given by \cite{don,jpa}:
\begin{equation}
\ddot{\bf r}_k=\Omega\, {\bf \hat z}\times \dot{\bf r}_k-\Omega\,\omega_k{\bf r}_k,\label{1}
\end{equation}
with
\begin{equation}
\Omega=\rho_s\kappa/m_v\label{2}
\end{equation}
and
\begin{equation}
\omega_k=\frac{\kappa k^2}{4\pi}[-\ln(|k|a)+0.116],\label{3}
\end{equation}
where $m_v$ denotes the vortex mass per unit length
and $\kappa=h/m_4$ denotes the quantum of circulation, given by the ratio of 
Planck's constant and the mass of one $^4$He atom.
The vortex core parameter $a\sim$ 1 \AA $\,$ in (\ref{3}) is assumed to be much less than the 
wavelength ($|k|a\ll 1$), which in turn ensures that $\omega_k\ll\Omega$ \cite{don,jpa}. The first
term in equation (\ref{1}) corresponds to the Magnus force, whereas the second term stems from the induced
velocity on the vortex line element by the helix curvature. Assuming that the vortex has a 
counterclockwise circulation, the frequencies (\ref{2}) and (\ref{3}) are positive, and then
equation (\ref{1}) turns out to be analogous to that ruling the two-dimensional motion of a 
negative point charge, in a uniform magnetic field parallel to the $z$ axis and subjected to a 
harmonic central force.
Such an equation derives from the following Hamiltonian:
\begin{equation}
\frac{m_v}{2}({\bf v}_k^2+\Omega\,\omega_k\,{\bf r}_k^2)\label{4},
\end{equation}
being
\begin{equation}
{\bf v}_k=\frac{{\bf p}_k}{m_v}+\frac{\Omega}{2}\,{\bf \hat z}\times {\bf r}_k,
\label{5}
\end{equation}
where ${\bf p}_k$ denotes the conjugate momentum to ${\bf r}_k$. In fact, from 
Hamilton equations it is easy to check that (\ref{5}) corresponds to the velocity 
$\dot{\bf r}_k$, while the acceleration is indeed given by (\ref{1}). From the above 
electromagnetic analogy it is also useful to represent the coordinate ${\bf r}_k$ as the sum
of the center coordinate ${\bf R}^{(k)}_0$ of the cyclotron circle plus the relative coordinate 
${\bf R}'_k$
from such a center \cite{yosh},
\begin{equation}
{\bf r}_k={\bf R}^{(k)}_0+{\bf R}'_k={\bf R}^{(k)}_0+
{\bf v}_k\times {\bf \hat z}/\Omega.
\label{6}
\end{equation}
%\begin{equation}
%{\bf R}'_k= {\bf v}_k\times {\bf \hat z}/\Omega.
%\label{7}
%\end{equation}
Then, from Hamilton equations one easily obtains the following pair of coupled equations for 
${\bf R}^{(k)}_0$ and ${\bf R}'_k$:
\begin{subequations}
\label{8}
\begin{eqnarray}
\dot{\bf R}^{(k)}_0 & = & -\omega_k\, {\bf \hat z}\times {\bf R}^{(k)}_0-\omega_k\, {\bf \hat z}
\times {\bf R}'_k\label{8a} \\
\dot{\bf R}'_k & = & (\Omega+\omega_k)\, {\bf \hat z}\times {\bf R}'_k+\omega_k\, {\bf \hat z}
\times {\bf R}^{(k)}_0\label{8b}.
\end{eqnarray}
\end{subequations}
The solution of the above system
 can more simply  be expressed in complex notation (${\bf r}=x+iy$), and to
first order in $\omega_k/\Omega$ we have,
\begin{subequations}
\label{9}
\begin{eqnarray}
{\bf R}^{(k)}_0 & = & A_1 \exp(-i\omega_k t)
-\frac{\omega_k}{\Omega}A_2 \exp[i\Omega
(1+\omega_k/\Omega) t]\label{9a} \\
{\bf R}'_k & = & A_2 \exp[i\Omega(1+\omega_k/\Omega) t]
-\frac{\omega_k}{\Omega}A_1 \exp(-i\omega_k t),\label{9b}
\end{eqnarray}
\end{subequations}
where $A_1$ and $A_2$ are complex numbers depending on initial conditions. Then, neglecting 
corrections of order $\omega_k/\Omega$, the solution corresponds to a decoupling of the system
(\ref{8}):
\begin{subequations}
\label{10}
\begin{eqnarray}
\dot{\bf R}^{(k)}_0 & = & -\omega_k\, {\bf \hat z}\times {\bf R}^{(k)}_0 \label{10a}\\
\dot{\bf R}'_k & = & \Omega\, {\bf \hat z}\times {\bf R}'_k,\label{10b}
\end{eqnarray}
\end{subequations}
that is, we shall simply have circular trajectories, namely for ${\bf R}^{(k)}_0$ a clockwise one
with angular frequency $\omega_k$ and for ${\bf R}'_k$ a 
counterclockwise one with angular frequency $\Omega$. Then, it becomes clear that ${\bf R}^{(k)}_0$
should be ascribed to Kelvin modes, while ${\bf R}'_k$ should correspond to the cyclotron 
ones \cite{don,jpa}.
Theoretical estimates of the vortex mass \cite{don,duan}
 lead to
cyclotron frequency values in (\ref{2}) of the order or greater than $k_BT/\hbar$ for 
$T< 1$ K. This seems to indicate that quantum effects could be of importance. Quantization
of coordinate and momentum arises straightforwardly from the electromagnetic analogy
and reads \cite{yosh},
\begin{eqnarray}
{\bf R}^{(k)}_0 & = & \sqrt{\frac{\hbar}{2\rho_s\kappa L}}\,[e^{-ikz}\beta^\dagger_k({\bf \hat x}+i{\bf \hat y})
+e^{ikz}\beta_k({\bf \hat x}-i{\bf \hat y})]\label{11}\\
{\bf R}'_k & = & \sqrt{\frac{\hbar}{2\rho_s\kappa L}}\,[e^{-ikz}\alpha_k({\bf \hat x}+i{\bf \hat y})
+e^{ikz}\alpha^\dagger_k({\bf \hat x}-i{\bf \hat y})]\label{12}\\
{\bf p}_k & = & i\sqrt{\frac{\hbar\rho_s\kappa}{8 L}}\,[e^{-ikz}(\beta^\dagger_k-\alpha_k)
({\bf \hat x}+i{\bf \hat y})
+e^{ikz}(\alpha^\dagger_k-\beta_k)({\bf \hat x}-i{\bf \hat y})],\label{13}
\end{eqnarray}
where $\alpha^\dagger_k$ ($\beta^\dagger_k$) denotes a creation operator of right (left) circular
quanta. The $z$ dependence of coordinates and momentum 
corresponds to the
rotation generated on following the helix path. 
Normal modes corresponding to $k{\bf \hat z}$
 are ruled by a
 Hamiltonian of the
form (\ref{4}):
\begin{equation}
H_k=\int_0^L dz\,\frac{m_v}{2}[{\bf v}_k^2+\Omega\,\omega_k\,{\bf r}^2_k]\label{14}
\end{equation}
where, taking into account (\ref{11}) to (\ref{13}), one may verify that ${\bf v}^2_k$ and 
${\bf r}^2_k$ do not 
depend on $z$ and also that the limit $\omega_k/\Omega\ll 1$ yields \cite{jpa},
\begin{equation}
H_k=\hbar\Omega(\alpha^\dagger_k\alpha_k+\frac{1}{2})+
\hbar\omega_k(\beta^\dagger_k\beta_k+\frac{1}{2})\label{15}.
\end{equation}
That is, both polarizations
(cyclotron and Kelvin modes) become decoupled, as seen from a classical viewpoint
in equations (\ref{10}). Here it is instructive to evaluate the limit of a massless vortex,
which is often found in the literature \cite{fet1}. In fact, the limit $\Omega\rightarrow
\infty$ in (\ref{1}) transforms such an equation into a first-order one, with the 
consequence that the Cartesian components $x_k$ and $y_k$ of ${\bf r}_k$ become conjugate variables.
Note that this amounts to ignoring the cyclotron motion by setting ${\bf r}_k\equiv{\bf R}^{(k)}_0$
(cf. Eq. (\ref{6})),
where the components of ${\bf R}^{(k)}_0$ in (\ref{11}) obey canonical
commutation relations.

The Hamiltonian of the $k=0$ modes reads \cite{jpa1}
\begin{equation}
H_0= \frac{m_v L}{2}(v_{0x}^2+v_{0y}^2)-\Omega_{\rm rot}\,\rho_s\kappa L y_0^2
\label{z1}
\end{equation}
with $v_{0x}=p_{0x}/m_v-\Omega y_0$, $v_{0y}=p_{0y}/m_v$.
The quantization of the coordinate ${\bf r}_0={\bf R}^{(0)}_0+{\bf R}'_0={\bf R}^{(0)}_0+
{\bf v}_0\times {\bf \hat z}/\Omega$  
remains given through
the previous expressions (\ref{11}) and (\ref{12}), whereas the momentum expression (\ref{13})
 becomes changed according to the
Landau gauge as ${\bf p}_0  =  \sqrt{\hbar\rho_s\kappa/(2 L)}\,[\,i(\beta^\dagger_0-\beta_0)
{\bf \hat x}+(\alpha^\dagger_0+\alpha_0){\bf \hat y}\,]$. The Hamiltonian (\ref{z1}) can be exactly
solved \cite{yosh} yielding in the limit $\Omega_{\rm rot}
\ll\Omega$ the
decoupling of cyclotron and translational modes:
\begin{equation}
H_0=\hbar\Omega(\alpha^\dagger_0\alpha_0+\frac{1}{2})+
\frac{\hbar\Omega_{\rm rot}}{2}(\beta^\dagger_0-\beta_0)^2.
\label{16}
\end{equation}
Thus, the Heisenberg equation of motion for the cyclotron coordinate 
${\bf R}'_0$ is given again by (\ref{10b}), whereas the translational
coordinate evolves according to the superfluid velocity field 
\begin{equation}
\dot{{\bf R}}^{(0)}_0  = -2\,\Omega_{\rm rot}\,Y^{(0)}_0\,{\bf \hat x}.
\label{p22a}
\end{equation}

In conclusion, the vortex Hamiltonian may be written 
as follows:
\begin{equation}
H_v=\sum_k\left[\hbar\Omega\left(\alpha^\dagger_k\alpha_k+\frac{1}{2}\right)+
\hbar\omega_k\left(\beta^\dagger_k\beta_k+\frac{1}{2}\right)\right]
+
\frac{\hbar\Omega_{\rm rot}}{2}(\beta^\dagger_0-\beta_0)^2
\label{18}
\end{equation}
with $k=2\pi m/L$ and $m$ an integer. 

\subsection{Quasiparticle Hamiltonian}\label{sec1B}
The normal fluid will be represented by the following Hamiltonian:
\begin{equation}
H_n=\sum_{ {\bf q}}\,\hbar\omega_{ {\bf q}} \,\,
a_{{\bf q}}^\dagger \,  a_{ {\bf q}}
\label{19},
\end{equation}
where $a_{{\bf q}}^\dagger$ denotes a creation operator of quasiparticle excitations 
of pseudomomentum $\hbar{\bf q}$ and frequency $\omega_{ {\bf q}}$. Such a frequency corresponds
to the lab frame and may be written as a Doppler-shifted frequency from the superfluid frame,
$\omega_{ {\bf q}}=\omega_q+{\bf q}\cdot{\bf v}_s$, where $\omega_q$ is the familiar (isotropic)
dispersion relationship of $^4$He quasiparticle excitations. Now, the background superflow velocity
should be much less than the Landau critical velocity $\sim$60 m/s, so we may safely 
approximate $\omega_{ {\bf q}}=\omega_q$ in (\ref{19}).
 Note also that we disregard any 
interaction between the 
quasiparticles themselves, since we
shall work at low enough temperature, so that they remain
dilute allowing their treatment	 as a noninteracting gas.

\subsection{Interaction Hamiltonian}\label{sec1C}
The interaction Hamiltonian between the vortex and the quasiparticles will be represented by the
pseudomomentum-conserving
form:
\begin{equation}
H_{\rm int}=\sum_{ {\bf p},  {\bf q} }\int_0^L dz
  \, \, \Lambda_{{\bf p}  {\bf q}} 
\,a_{{\bf p}}^\dagger \,  a_{ {\bf q}} 
\exp[-i({\bf p}-{\bf q})\cdot{\bf r}(z)
-i(p_z-q_z)z]\label{20},
\end{equation}
where ${\bf r}(z)={\bf R}_0(z)+{\bf R}'(z)$, being
${\bf R}_0(z)=\sum_k{\bf R}_0^{(k)}(z)$ and ${\bf R}'(z)=\sum_k{\bf R}'_k(z)$,
with ${\bf R}_0^{(k)}(z)$ and 
${\bf R}'_k(z)$ given by
(\ref{11}) and (\ref{12}), respectively. The parameters $\Lambda_{{\bf p}
  {\bf q}}$ in (\ref{20})
  represent
scattering amplitudes depending on wave vectors of scattered 
quasiparticles. 
The vortex pseudomomentum per unit length \cite{jpa1}
${\bf K}(z)=-\rho_s\kappa{\bf \hat z}\times{\bf R}_0(z)$ integrated along the vortex line 
yields the generator of vortex translations or vortex pseudomomentum $\int_0^L
{\bf K}(z)dz=-\rho_s\kappa L\,{\bf \hat z}\times{\bf R}_0^{(0)}$, which involves only
translational coordinates, as expected. 
Then, adding
such a pseudomomentum to the quasiparticle pseudomomentum $\sum_{ {\bf q}}\,\hbar{\bf q} \,\,
a_{{\bf q}}^\dagger \,  a_{ {\bf q}}$, we have the pseudomomentum of the whole system, which
can be shown to commute with $H_{\rm int}$.
Note that only the $x$-component of the vortex pseudomomentum
$\rho_s\kappa L\,Y_0^{(0)}$
will commute with the vortex Hamiltonian (\ref{18}), unless $\Omega_{\rm rot}=0$. 
This result may be easily interpreted, since
a superflow of velocity ${\bf v}_s=-2\,\Omega_{\rm rot}\,y\,{\bf \hat x}$ 
produces a translation symmetry breaking in
the $y$-direction.

\section{Equation of motion for the normal fluid flow}\label{s3}
The interaction Hamiltonian (\ref{20}) is difficult to deal with, 
so recalling the low 
amplitude of the helical oscillations, we may 
rewrite the exponential in 
(\ref{20}) as 
\begin{eqnarray}
& &\exp[-i({\bf p}-{\bf q})\cdot{\bf r}(z)-i(p_z-q_z)z] = \nonumber \\
& &
\exp[-i({\bf p}-{\bf q})\cdot{\bf R}_0^{(0)}]\,\exp[-i(p_z-q_z)z]\,
\exp[-i({\bf p}-{\bf q})\cdot({\bf r}(z)-{\bf R}_0^{(0)}
)]
\label{eqq}
\end{eqnarray}
 and next approximate the last exponential to first 
order in ${\bf r}(z)-{\bf R}_0^{(0)}$. 
This procedure, however, is not valid for vortex modes with frequencies
approaching zero, i.e. the lowest part of Kelvin's spectrum, as noted early by 
Fetter \cite{fet2}. In fact, he showed that retaining a finite number of terms of
such an
exponential expansion leads to divergent results, analogous to those
of the ``infrared catastrophe'' in electrodynamics.
Here it is expedient to recall that 
 within our study, each 
vortex forms part of a vortex lattice which will be regarded as a heat bath 
for the quasiparticle
flow. Now, it is well known that rather simple models of heat bath often provide suitable 
descriptions of realistic
environments \cite{cal}. 
Relying on this hypothesis and to overcome the above difficulty, we shall represent Kelvin's spectrum
by a single frequency $w_0$, which will be eventually regarded as a temperature-dependent parameter
in order to take into account the distinct features of 
the interaction of such waves with phonons and rotons. 
In summary, we shall make use of a simplified model of heat bath consisting of vortex modes of two 
frequencies ($w_0\ll\Omega$) of opposite polarization.
On the other hand, neglecting the vortex displacement in the 
$y$-direction \cite{jpa1},
we shall replace the translational coordinate operator in the first
 exponential on the right-hand side of Eq. (\ref{eqq}) by the $c$-number
$\langle{\bf R}_0^{(0)}\rangle=-2\,\Omega_{\rm rot}\,\langle Y^{(0)}_0\rangle t\,{\bf \hat x}$
(cf. Eq. (\ref{p22a})). Such a replacement
becomes equivalent to having  time-dependent
 scattering amplitudes  in  Eq. (\ref{20}), i.e. with a time dependent phase 
factor, $\Lambda_{{\bf p}  {\bf q}}
\,\exp[i(p_x-q_x)2\,\Omega_{\rm rot}\,\langle Y^{(0)}_0\rangle t]$. This factor, however,
would not have any
practical incidence, since our results will be shown to be dependent upon the absolute 
value of the scattering
amplitudes.
 Finally, taking into account these approximations
we may replace  $\exp[-i({\bf p}-{\bf q})\cdot{\bf r}(z)]\simeq
1-i({\bf p}-{\bf q})\cdot({\bf r}(z)-{\bf R}_0^{(0)})$
in Eq. (\ref{20}) yielding \cite{jpa}
\begin{eqnarray}
H_{\rm int} & = & \sqrt{\frac{\hbar L}{2\rho_s\kappa}}\,
\sum_{k, {\bf p},  {\bf q} }\delta_{p_zq_z} \Lambda_{{\bf p}  {\bf q}}
\,a_{{\bf p}}^\dagger \,  a_{ {\bf q}} \;
\{[(q_y-p_y)
+i(q_x
-p_x)][\alpha_k^\dagger+(1-\delta_{k0})\beta_k]\nonumber \\
& + & [(p_y-q_y)
+i(q_x-p_x)][\alpha_k+(1-\delta_{k0})\beta_k^\dagger]\}.\label{21}
\end{eqnarray}

There is an additional parameter to be taken into account in our vortex heat bath, 
that is the total number of modes
$2L/\lambda_{\rm min}$, 
where the factor 2 in front of the expression comes from the two possible signs for $k$.
We shall assume for simplicity  that both polarizations have a common 
`ultraviolet' cutoff $\lambda_{\rm min}$, which according to Sec. \ref{sec1A},
should be greater than the vortex 
core parameter ($\sim$ 1 \AA) and the mean radius of the helix.
Such a radius turns out to be of the order of the core parameter for cyclotron modes, while
for Kelvin modes in a lattice of $\Omega_{\rm rot}\sim$ 1 s$^{-1}$ has been 
estimated \cite{fet1} as
$\sim$ 10$^3$ \AA$\,\sqrt{T/\rm K}$. So we shall assume $\lambda_{\rm min}\sim 10^3$ \AA~
in our calculations.

A total Hamiltonian given by the sum of (\ref{18}), (\ref{19}) and (\ref{21}) yields
 dissipative evolutions for each vortex mode, which for cyclotron modes are
ruled by \cite{jpa}:
\begin{equation}
\langle \ddot{\bf R}'_k\rangle=\Omega\left(1-\frac{D'}{\rho_s\kappa}\right)
\, {\bf \hat z}\times \langle\dot{\bf R}'_k\rangle
-\Omega\,\frac{D}{\rho_s\kappa}\langle\dot{\bf R}'_k\rangle,\label{27}
\end{equation}
where explicit expressions for the longitudinal and transverse friction coefficients,
$D$ and $D'$, respectively, can be found in \cite{jltp}.
There we have shown that a scattering amplitude given by \cite{prb}
\begin{equation}
\Lambda_{{\bf p}  {\bf q}}=\frac{\hbar\kappa}{Vc_s}\sqrt{\frac{19}{140}|\omega'_p|
|\omega'_q|},
\label{22}
\end{equation}
where $V$
 denotes the volume of the system and $\omega'_p$ ($c_s$) denotes the quasiparticle group
(sound) velocity,
leads to
a very good agreement with the experimental determinations of the longitudinal friction
coefficient for temperatures below 1 K. The presence of the viscous force represented 
by the last term in (\ref{27}) leads to a vanishing asymptotic velocity, 
$\langle\dot{\bf R}'_k\rangle_{t\rightarrow\infty}=0$, and the same behavior will 
present the  coordinate, $\langle{\bf R}'_k\rangle_{t\rightarrow\infty}=0$.

To summarize, we have focused in previous works on the  dynamics of an individual vortex
line which interacts with a quasiparticle heat bath. Now, being focused on 
the time evolution of the normal fluid, we shall concentrate on
the dissipative
dynamics of a quasiparticle flow which interacts with the heat bath formed by a uniform
array of $N_v$ quantized vortex filaments.
We have derived in Appendix \ref{ap1} the system (\ref{a6}) of
non-Markovian equations that rule the time evolution of the quasiparticle populations $n_{\bf q}$.
Then, from (\ref{a6}) one easily obtains the following equation of motion for the quasiparticle 
pseudomomentum:  
\begin{eqnarray}
\sum_{ {\bf q}}\,\hbar{\bf q}\,\dot{n}_{\bf q} & = & \frac{2L^2N_v}{\rho_s\kappa\lambda_{\rm min}}
\sum_{ {\bf p},\;  {\bf q},\;i }  \, \, |\Lambda_{{\bf p}  {\bf q}}|^2
\,\delta_{p_zq_z} ({\bf p}-{\bf q})({\bf p}-{\bf q})^2
\int_0^t d\tau \cos[(\omega_p-\omega_q+ w_i)\tau]
 \{n_{\bf q}(t-\tau)\nonumber\\
& \times & [1
+n_{\bf p}(t-\tau)]
+[n_{\bf q}(t-\tau)-n_{\bf p}(t-\tau)]
\,[e^{\hbar w_i/k_BT}-1]^{-1}\},
\label{30}
\end{eqnarray}
where $w_i$ denotes the frequencies $\Omega$ and $w_0$.
We shall assume that the
quasiparticle numbers in the above expression are well described by a local equilibrium form:
\begin{equation}
n_{\bf q}=[e^{\hbar\omega_{\bf q}/k_BT}-1]^{-1}
\simeq[e^{\hbar\omega_q/k_BT}-1]^{-1}
+
\frac{\hbar\,{\bf q}\cdot[{\bf v}_n-{\bf v}_s]}{4k_BT\sinh^2(\hbar\omega_q/2k_BT)},
\label{31}
\end{equation}
where the quasiparticle frequency is measured from a reference frame where 
the local normal fluid is at rest,
$\omega_{\bf q}=\omega_q-{\bf q}\cdot[{\bf v}_n-{\bf v}_s]$.
Thus, the quasiparticle 
pseudomomentum $\sum_{ {\bf q}}\,\hbar{\bf q}\,n_{\bf q}
= A L \rho_n ({\bf v}_n-{\bf v}_s)$ corresponds to 
a macroscopically small area $A$ of the $x$-$y$ plane
containing $N_v$ vortices, where the spatial dependence of the
fields ${\bf v}_n$ and ${\bf v}_s$ can be neglected.
Then, assuming a time dependence stemming exclusively from ${\bf v}_n(t)$ (cf. Sec. \ref{s1}),
a straightforward calculation leads to the following
non-Markovian equation:
\begin{equation}
\dot{\bf v}_n = -\int_0^td\tau\,[{\bf v}_n(t-\tau)-{\bf v}_s]\,\mu(\tau),
\label{32}
\end{equation}
with a memory kernel given by
\begin{eqnarray}
\mu(\tau)
 & = & \frac{L\,\hbar\,\Omega_{\rm rot}}{k_BT\rho_n\rho_s\kappa^2\lambda_{\rm min}}
\sum_{ {\bf p},\;  {\bf q},\;i }  \, \, |\Lambda_{{\bf p}  {\bf q}}|^2
\,\delta_{p_zq_z}\cos[(\omega_p-\omega_q
+ w_i)\tau]\,
\{ |{\bf p}-{\bf q}|^4\,g_+(w_i)\nonumber\\
& + & [|{\bf \hat z}\times({\bf p}\times{\bf \hat z})|^4
-|{\bf \hat z}\times({\bf q}\times
{\bf \hat z})|^4]\,g_-(w_i)\}\,
n(\omega_q)\,[1+n(\omega_p)]\,[1+n( w_i)],
\label{33}
\end{eqnarray}
where
\begin{equation}
g_\pm(w_i) = \frac{n(\omega_p)}{n(\omega_q- w_i)}\pm 
\frac{n(\omega_q)\exp[\hbar(\omega_q-\omega_p- w_i)/k_BT]}{n(\omega_p+ w_i)}
\label{4.8}
\end{equation}
and $n(w)=[\exp(\hbar w/k_BT)-1]^{-1}$. In the thermodynamic limit, the summations 
over ${\bf p}$ and ${\bf q}$ in
(\ref{33}) become integrals and $\mu(\tau)$ acquires a finite lifetime. If such a lifetime
can be regarded as microscopic in comparison with the observational timescale,
equation (\ref{32}) may be transformed according to the Markov approximation 
into the differential equation:
\begin{equation}
\dot{\bf v}_n=-\nu[{\bf v}_n(t)-{\bf v}_s]
\label{35}
\end{equation}
with
\begin{eqnarray}
\nu & = & \int_0^\infty\mu(\tau)\,d\tau 
 =  \frac{L\,h\,\Omega_{\rm rot}}{k_BT\rho_n\rho_s\kappa^2\lambda_{\rm min}}
\sum_{ {\bf p},\;  {\bf q},\;i }   |\Lambda_{{\bf p}  {\bf q}}|^2
\,\delta_{p_zq_z}
\delta(\omega_p-\omega_q+ w_i)\nonumber \\
& \times &|{\bf p}-{\bf q}|^4 
 \,
n(\omega_q)\,[1+n(\omega_p)]\,[1+n( w_i)],
\label{36}
\end{eqnarray}
where the continuum limit corresponds to the replacement
$\sum_{ {\bf p} , {\bf q} }  \, \, \delta_{p_zq_z}\rightarrow
[A^2L/(2\pi)^5]\int d^3{\bf p}
\int d^3{\bf q}\,\,\, \delta(p_z-q_z)$.
Actually, we have studied in Appendix \ref{ap2} the non-Markovian equation (\ref{32})
finding that memory effects are  negligible for $\Omega_{\rm rot}\ll w_0$, which will
be assumed hereafter.
Finally, taking into account (\ref{35}),
(\ref{p1})
and (\ref{p3}), we may obtain the expression of the mutual friction parameter from $B=\nu/\Omega_{\rm rot}$.

\section{Study of the mutual friction parameter $B$}\label{s4}
An explicit expression for the friction parameter $B$ can be extracted by computing
the right-hand side of equation (\ref{36}) (see Appendix \ref{ap2}):
\begin{equation}
B  =  \frac{19 \hbar^3}{70(2\pi)^2
c_s^2\rho_n\rho_sk_BT\lambda_{\rm min}}\sum_i\,[1+n(w_i)]
\,\int_0^\infty dp\,|\omega'_p|
n(\omega_p+w_i)\,[1+n(\omega_p)]
\sum_j \Gamma(p,q_j^{(i)}),
\label{37}
\end{equation}
where
\begin{equation}
 \Gamma(p,q)=   \left\{
\begin{array}{r}
              p^2q(q^4+p^4/5+2p^2q^2)\,\,\,\,\,\,(p\leq q) \\
              q^2p\,(p^4+q^4/5+2p^2q^2)\,\,\,\,\,\,(q\leq p)
            \end{array}
\right.
\label{38}
\end{equation}
and $q=q_j^{(i)}$ denote the roots of the equation $\omega_q=\omega_p+w_i$.
From (\ref{37}) we may see that $B$ consists of two terms arising from the frequencies
$w_i=w_0$ and $w_i=\Omega$. Such contributions, 
however, are weighted by respective factors $[1+n(w_0)]
\gg [1+n(\Omega)]$,
so the cyclotron contribution will be always negligible
with respect to that arising from the frequency $w_0$
and we have that, in practice, $B$ will correspond to the limit of massless vortices, $\Omega
\rightarrow\infty$.
The expression (\ref{37}) leads to simpler phonon and roton approximations. 
If we restrict ourselves to  
$w_0\lesssim$ 10$^{10}$ s$^{-1}$, such a frequency may be neglected
everywhere in (\ref{37}), except in the factor $[1+n(w_0)]$. 
Then, to approximate for 
phonon temperatures (T$<$0.4 K), we use the linear dispersion relation $\omega_p
=c_sp$ and get
\begin{equation}
B_{\rm ph}=\frac{254.9\,[1+n(w_0)]\,(k_BT)^3}{\hbar^2c_s^4\rho_s\lambda_{\rm min}}.
\label{39}
\end{equation}
On the other hand, for temperatures above 0.6 K, only the portion of the dispersion curve
around the roton minimum makes a significant contribution to the integrand in (\ref{37}),
then making use of the usual approximations in roton calculations \cite{prb}, we obtain
\begin{equation}
B_{\rm r}=\frac{2.079\,[1+n(w_0)]\,\hbar k_0^3\sqrt{k_BT}}{c_s^2\sqrt{\mu}
\rho_s\lambda_{\rm min}},
%\simeq 0.22\,[1+n(w_0)]\,\sqrt{T}\,
%{\rm  K}^{-\frac{1}{2}},
\label{40}
\end{equation}
where $\mu$ and $k_0$ are parameters entering the
 Landau para\-bolic approximation, $\omega_p=\Delta/\hbar+
\hbar(p-k_0)^2/2\mu$.

Experimental determinations of $B$ have been reported only above 1.3 K.
However, we may utilize the following expression valid
 for temperatures below 1 K \cite{don},
\begin{equation}
B=\frac{2 D}{\rho_n\kappa}
\label{41p}
\end{equation}
and replace $D$ in (\ref{41p}) 
by means of the Iordanskii theoretical estimate for the phonon temperature range \cite{ior}, yielding
\begin{equation}
B_{\rm ph}=8.17\,\,\frac{k_BT}{m_4c_s^2}.
\label{41pp}
\end{equation}
On the other hand, for the roton temperature range, we may replace $D=\rho_nv_G\sigma_
\parallel$ in (\ref{41p}) yielding,
\begin{equation}
B_{\rm r}=\frac{2\sigma_
\parallel}{\kappa}\sqrt{\frac{2k_BT}{\pi\mu}}\simeq 1.5\sqrt{T}\,\,{\rm  K}^{-\frac{1}{2}},
\label{41}
\end{equation}
being $v_G$ the average group velocity of rotons and  $\sigma_\parallel\simeq 8.38$ \AA~the 
roton scattering length \cite{don,bar}. Notice that we have replaced the 
friction coefficient $D$ by expressions
corresponding to a straight vortex, since corrections due to vortex
bending should be negligible, as seen in Sec. \ref{s1}.
 This of course does not mean that vortices
would remain straight against the quasiparticle scattering; on the contrary, thermal excitation
of Kelvin waves is undoubtedly expected to occur, although details of this process
are not evident from expressions (\ref{41pp}) and (\ref{41}). 
However, some features about such a process may be deduced from our 
results (\ref{39}) and (\ref{40}). 
First it is convenient to discuss the physical meaning of the frequency $w_0$.
Since such a frequency is intended for representing the whole spectrum 
of Kelvin waves (\ref{3}) in the context of an interaction with quasiparticles,
 it should not be surprising to find that quite distinct values of $w_0$
could be required in order to better estimate interactions with phonons
or rotons. This amounts to assuming a dependence of $w_0$ on temperature, which may be fully extracted by
equating our results with the theoretical expressions (\ref{41pp}) and (\ref{41}).
In fact, from (\ref{39}) and (\ref{41pp}) we may conclude that phonon scattering
at a temperature $T$ should be expected to excite Kelvin waves about a representative frequency given by
\begin{equation}
w_0\simeq 5.64\times 10^{8}{\rm s}^{-1}{\rm K}^{-3}\,T^3,
\label{phono}
\end{equation}
while (\ref{40}) and (\ref{41}) imply that roton scattering should be likely to excite Kelvin waves 
about frequency
\begin{equation}
w_0\simeq 2.08\times 10^{10}{\rm s}^{-1}{\rm K}^{-1}\,T.
\label{roto}
\end{equation}
Recall that according to the Markov approximation (Sec.~\ref{s3}) one should assume 
$w_0\gg\Omega_{\rm rot}\sim$ 1 s$^{-1}$, which sets up 
a lower bound for the validity of the result (\ref{phono})
at temperatures above $\sim 0.01$ K. In addition, the assumption 
$\lambda_{\rm min}\sim 10^3$ \AA~implies an upper bound for the Kelvin spectrum,
$\max(\omega_k)\sim  10^{8}\,{\rm s}^{-1}$, which turns out to be consistent with
a phonon temperature range below 0.4 K in (\ref{phono}).
However, the values arising from (\ref{roto})
seem to be overestimated for roton temperatures, since they would only be consistent with
a $\lambda_{\rm min}$ of order $
10^2$ \AA. In addition, such values of $\omega_0$ could reach the
order of cyclotron frequencies arising from some theories of the
vortex mass \cite{duan}, contradicting the assumption $\omega_k\ll\Omega$ of Sec.~\ref{sec1A}. 
This suggests that only the qualitative trend $w_0\sim T$ should be taken into account from the 
result (\ref{roto}).

\section{Concluding remarks}\label{s7}
We have analyzed a microscopic model of mutual friction represented by the
dissipative dynamics of a normal fluid flow, which interacts with the helical normal modes of
vortices comprising a lattice in thermal equilibrium.
Such vortices interact with the quasiparticles forming
 the normal fluid through
a pseudomomentum-conserving scattering Hamiltonian.
Assuming a simplified model for the vortex heat bath,
we have derived an equation of motion for the quasiparticle pseudomomentum leading to
the expected form predicted by the HVBK equations. 
We have shown that the mutual friction coefficient $B$ turns out to be practically independent of
the values of vortex mass arising from diverse theories. 
Finally, from a comparison of our expression of $B$ with previous theoretical estimates,
we have deduced interesting qualitative
features about the interaction of quasiparticles with
Kelvin modes, namely phonon (roton) scattering at a temperature $T$ should be expected to excite Kelvin
waves about representative frequencies proportional to $T^3$
($T$).

\section*{Acknowledgments}
This work has been performed under Grant PIP 5409 from CONICET, Argentina.

\appendix

\section{Derivation of the equation of motion for the quasiparticle populations}\label{ap1}

Our starting point is the Heisenberg equation for the quasiparticle number operator
$a_{{\bf q}}^\dagger \,  a_{ {\bf q}}$:
\begin{eqnarray}
\frac{d}{dt}(a_{{\bf q}}^\dagger \,  a_{ {\bf q}}) & = & 
\sqrt{\frac{L}{2\hbar \rho_s\kappa}}\,
\sum_{k, {\bf p},j}\delta_{p_zq_z} \{\Lambda_{{\bf p}  {\bf q}}\,
[(p_x-q_x)+i(q_y
-p_y)]
\,a_{{\bf p}}^\dagger \,  a_{ {\bf q}} \;[\alpha_k^{(j)\dagger}+(1-\delta_{k0})\beta_k^{(j)}] 
\nonumber\\ & + &  \Lambda^*_{{\bf p}  {\bf q}}\,
[(p_x-q_x)
-i(q_y-p_y)]\,
[\alpha_k^{(j)}+(1-\delta_{k0})
\beta_k^{(j)\dagger}]\,a_{{\bf q}}^\dagger \,  a_{ {\bf p}}
+\Lambda_{{\bf p}  {\bf q}}\,
[(p_x - q_x)
-i(q_y-p_y)]
a_{{\bf p}}^\dagger \,  a_{ {\bf q}} \;[\alpha_k^{(j)}
\nonumber\\ & + &
(1-\delta_{k0})\beta_k^{(j)\dagger}] 
+\Lambda^*_{{\bf p}  {\bf q}}\,
[(p_x-q_x) 
 +  i(q_y-p_y)]\,
[\alpha_k^{(j)\dagger}+(1-\delta_{k0})\beta_k^{(j)}]\,a_{{\bf q}}^\dagger \,  a_{ {\bf p}}
\},\label{a1}
\end{eqnarray}
where $j$ labels each vortex ($1\leq j \leq N_v$) and
we have taken into account that only the interaction Hamiltonian 
(\ref{21}) yields a nonvanishing commutator with such an operator. 
Next we 
write out the following expressions for the operators appearing on the right-hand
side of the above equation, which arise from the formal solutions of the
corresponding Heisenberg equations:
\begin{eqnarray}
a_{{\bf p}}^\dagger(t) \,  a_{ {\bf q}}(t) & = &
e^{i(\omega_p-\omega_q)t}\,a_{{\bf p}}^\dagger(0) \,  a_{ {\bf q}}(0)+
i\sqrt{\frac{L}{2\hbar \rho_s\kappa}}\,\int_0^t d\tau 
e^{i(\omega_p-\omega_q)\tau}\,
\sum_{k, {\bf q'},j}\delta_{p_zq'_z}\, \Lambda_{{\bf q'}  {\bf p}}\,
\{[(p_y-q'_y)
 + i(p_x-q'_x)] \nonumber\\
&\times&[\alpha_k^{(j)\dagger}(t-\tau)+(1-\delta_{k0})\beta_k^{(j)}(t-\tau)]
+ [(q'_y-p_y) 
+ i(p_x-q'_x)]\,[\alpha_k^{(j)}(t-\tau) 
 +(1-\delta_{k0})  \beta_k^{(j)\dagger}(t-\tau)]\}
\nonumber\\
&\times& a_{{\bf q'}}^\dagger(t-\tau) \,  a_{ {\bf q}}(t-\tau)
- \delta_{q_zq'_z}\, \Lambda_{{\bf q}  {\bf q'}}\,
\{[(q'_y-q_y) 
 +  i(q'_x-q_x)]\;[\alpha_k^{(j)\dagger}(t  
 -\tau)
+(1-\delta_{k0})\beta_k^{(j)}(t-\tau)]\nonumber\\
& + &
[(q_y-q'_y)+i(q'_x-q_x)]\;[\alpha_k^{(j)}(t-\tau) 
+ (1-\delta_{k0})\beta_k^{(j)\dagger}(t-\tau)]\} 
\,a_{{\bf p}}^\dagger(t-\tau) \,  a_{ {\bf q'}}(t-\tau),
\label{a2}
\end{eqnarray}
\begin{equation}
\alpha_k^{(j)\dagger}(t) = 
e^{i\Omega t}\,\alpha_k^{(j)\dagger}(0) +
\sqrt{\frac{L}{2\hbar \rho_s\kappa}}\,\int_0^t d\tau\,
e^{i\Omega\tau}
\sum_{{\bf p'},{\bf q'}}\delta_{p'_zq'_z}
 \Lambda_{{\bf p'}  {\bf q'}}\,
[(p'_x-q'_x) 
 + i(p'_y-q'_y)]\;a_{{\bf p'}}^\dagger(t-\tau)a_{ {\bf q'}}(t-\tau)
\label{a3}
\end{equation}
\begin{equation}
\beta_k^{(j)\dagger}(t) = 
e^{i w_0 t}\,\beta_k^{(j)\dagger}(0) +
\sqrt{\frac{L}{2\hbar \rho_s\kappa}}\,\int_0^t d\tau\,
e^{iw_0\tau}
\sum_{{\bf p'},{\bf q'}}\delta_{p'_zq'_z}\, \Lambda_{{\bf p'}  {\bf q'}}\,
[(p'_x-q'_x) 
 - i(p'_y-q'_y)]a_{{\bf p'}}^\dagger(t-\tau)   a_{ {\bf q'}}(t-\tau)
\label{a3p}
\end{equation}
The above expressions and their Hermitian conjugates are then replaced on
the right-hand side of (\ref{a1}), retaining only second-order terms in the
scattering amplitudes (weak-coupling approximation).
The resulting expression becomes greatly simplified when taking its ensemble
average according to the following prescriptions:

(i) All vortex-quasiparticle correlations are neglected, i.e. any average of a
product of vortex and quasiparticle operators is approximated by the corresponding
product of vortex and quasiparticle separate averages.
Such a procedure may be regarded as arising from, (a) an assumption of vanishing
initial correlations, i.e. assuming a whole system density operator given by
a product of a vortex operator and a quasiparticle operator, and (b) the above
weak-coupling approximation by which such vortex-quasiparticle averages should
be calculated to zeroth-order in the scattering amplitudes.

(ii) The assumption of a vortex heat bath corresponds to vortex operator averages
represented by thermal equilibrium expressions, i.e. 
$\langle\alpha_k^{(j)\dagger}\alpha_k^{(j)}\rangle=[\exp(\hbar\Omega/k_BT)-1]^{-1}$,
$\langle\beta_k^{(j)\dagger}\beta_k^{(j)}\rangle=[\exp(\hbar w_0/k_BT)-1]^{-1}$
and $\langle\alpha_k^{(j)\dagger}\rangle=\langle\beta_k^{(j)\dagger}\rangle=0$.

(iii) We assume that the system is close to equilibrium so that all nondiagonal
quasiparticle averages can be neglected, i.e.
\begin{equation}
\langle a_{{\bf p}}^\dagger \,  a_{ {\bf q}}\rangle=
\delta_{{\bf p}{\bf q}}n_{{\bf p}}.
\label{a5}
\end{equation}

Thus, we arrive at the following system of non-Markovian equations for the 
quasiparticle populations $n_{{\bf q}}$:
\begin{eqnarray}
\frac{dn_{\bf q}}{dt} & = & \frac{2L^2N_v}{\hbar\,\rho_s\kappa\lambda_{\rm min}}
\sum_{ {\bf p},i }  \, \, |\Lambda_{{\bf p}  {\bf q}}|^2
\,\delta_{p_zq_z} ({\bf p}-{\bf q})^2
\{\int_0^t d\tau \cos[(\omega_q-\omega_p+w_i)\tau]\,\{n_{\bf p}(t-\tau)
[1+n_{\bf q}(t-\tau)]\nonumber\\
&+&[n_{\bf p}(t-\tau)-n_{\bf q}(t-\tau)]\,[e^{\hbar w_i/k_BT}-1]^{-1}\}
\nonumber\\
&-&\int_0^t d\tau \cos[(\omega_p-\omega_q
+ w_i)\tau]\,
\{n_{\bf q}(t-\tau)
[1+n_{\bf p}(t-\tau)]+[n_{\bf q}(t-\tau)-n_{\bf p}(t-\tau)]
[e^{\hbar w_i/k_BT}-1]^{-1}\}\},
\label{a6}
\end{eqnarray}
where $w_i$ denotes the frequencies $\Omega$ and $w_0$.
It is easy to verify that (\ref{a6}) fulfills $\sum_{ {\bf q} } 
dn_{\bf q}/dt=0$, in agreement with the quasiparticle number-conserving Hamiltonian.
It is also instructive to rewrite (\ref{a6}) in the Markovian limit,
\begin{eqnarray}
\frac{dn_{\bf q}}{dt} & = & \frac{2\pi L^2N_v}{\hbar\,\rho_s\kappa\lambda_{\rm min}}
\sum_{ {\bf p},i }  \, \, |\Lambda_{{\bf p}  {\bf q}}|^2
\,\delta_{p_zq_z} ({\bf p}-{\bf q})^2
\{\delta(\omega_q-\omega_p+ w_i)\,\{n_{\bf p}(t)[1+n_{\bf q}(t)]
 +[n_{\bf p}(t)-n_{\bf q}(t)]\,[e^{\hbar w_i/k_BT}-1]^{-1}\}\nonumber\\
&-&
\delta(\omega_p-\omega_q+ w_i)\,
\{n_{\bf q}(t)
[1+n_{\bf p}(t)]
+ [n_{\bf q}(t)-n_{\bf p}(t)]
[e^{\hbar w_i/k_BT}-1]^{-1}\}\},
\label{a7}
\end{eqnarray}
and verify that thermal equilibrium populations, 
$n_{\bf p}=[e^{\hbar\omega_p/k_BT}-1]^{-1}$, lead to a vanishing result on the
right-hand side of the above equation.

\section{Study of memory effects}\label{ap2}
To explore the validity of the Markov approximation, it is convenient to
Laplace-transform equation (\ref{32}) according to $\tilde{\bf v}(z)=\int_0^\infty
e^{izt}[{\bf v}_n(t)-{\bf v}_s]dt$, (Im $z>0$). Then one easily finds that
\begin{equation}
\tilde{\bf v}(z)=[{\bf v}_n(0)-{\bf v}_s]/[\tilde{\mu}(z)-iz],
\label{b1}
\end{equation}
where the Laplace transform of the memory kernel can be written as a Cauchy
integral:
\begin{equation}
\tilde{\mu}(z)=\frac{1}{2\pi i}\int_{-\infty}^{+\infty}d\omega\,\,\frac{\nu(\omega)
+\nu(-\omega)}{\omega-z},
\label{b2}
\end{equation}
with,
\begin{eqnarray}
\nu(\omega)
 & = & \frac{h\, L\,\Omega_{\rm rot}}{2\,\lambda_{\rm min}k_BT\rho_n\rho_s\kappa^2}
\sum_{ {\bf p},\;  {\bf q},\;i }  \, \, |\Lambda_{{\bf p}  {\bf q}}|^2
\,\delta_{p_zq_z}\,\delta(\omega_p-\omega_q+ w_i+\omega)\,
\{ |{\bf p}-{\bf q}|^4\,g_+( w_i) +[|{\bf \hat z}\times({\bf p}\times{\bf \hat z})|^4
\nonumber\\
& - & 
|{\bf \hat z}\times({\bf q}\times{\bf \hat z})|^4]\,g_-( w_i)\}\,
n(\omega_q) [1+n(\omega_p)]\,[1+n( w_i)].	
\label{b3}
\end{eqnarray}
The poles of $\tilde{\bf v}(z)$ in (\ref{b1}) arise from the equation $z=-i\tilde{\mu}
(z)$ and the Markov approximation corresponds to the solution $z_M=-i\tilde{\mu}
(z\rightarrow i 0^+)=-i\nu$ [cf.  (\ref{36})]. A non-Markovian solution can be 
found iteratively, i.e. we may begin with the Markovian ansatz $z_0=-i\tilde{\mu}(0)$ 
and then proceed with $z_1=-i\tilde{\mu}(z_0)$, $z_2=-i\tilde{\mu}(z_1)$,
 and so on. Then the solution $z_s$ 
is better worked out in terms of the Taylor expansion of $\tilde{\mu}(z)$ around the origin:
\begin{equation}
z_s=-i\nu[1-i\tilde{\mu}'(0)-\tilde{\mu}''(0)\tilde{\mu}(0)/2+\cdots],
\label{b4}
\end{equation}
where the second and third terms inside the square brackets represent first- and second-order
corrections to the Markov approximation, respectively.
The Cauchy integral (\ref{b2}) and
its derivatives in equation (\ref{b4}) can be written as \cite{jpa}
\begin{subequations}
\label{b5}
\begin{eqnarray}
\tilde{\mu}(0)& = &\nu(0)\label{b5a}\\
\tilde{\mu}'(0) & = & \frac{1}{2\pi i}\int_{-\infty}^\infty \frac{d\omega}{\omega^2}
[\nu(\omega)+\nu(-\omega)-2\nu(0)]\label{b5b}\\
\tilde{\mu}''(0)& = &\nu''(0),
\label{b5c}
\end{eqnarray}
\end{subequations}
where the calculation of $\nu(\omega)$ from expression (\ref{b3})
may be reduced to a single one-dimensional integral (see \cite{prb}):
\begin{eqnarray}
\nu(\omega) & = & \frac{19 \hbar^3\,\Omega_{\rm rot}}{140(2\pi)^2
c_s^2\rho_n\rho_sk_BT\lambda_{\rm min}}\sum_i\, [1+n(w_i)]
\,\int_0^\infty dp |\omega'_p|\,n(\omega_p+w_i+\omega)\,[1+n(\omega_p)]
\nonumber\\ &\times& \sum_j \,\Gamma(p,q_j^{(i)})g_+(w_i)+\Phi(p,q_j^{(i)})g_-(w_i),
\label{b6}
\end{eqnarray}
where $g_\pm(w_i)$ is given by equation (\ref{4.8}) with
\begin{equation}
\omega_q=\omega_p+w_i+\omega,
\label{b7}
\end{equation}
\begin{equation}
 \Gamma(p,q)=   \left\{
\begin{array}{r}
              p^2q(q^4+p^4/5+2p^2q^2)\,\,\,\,\,\,(p\leq q) \\
              q^2p\,(p^4+q^4/5+2p^2q^2)\,\,\,\,\,\,(q\leq p)
            \end{array}
\right.
\label{b8}
\end{equation}
\begin{equation}
 \Phi(p,q)=   \left\{
\begin{array}{r}
              p^2q(-q^4+p^4/3+\frac{2}{3}p^2q^2)\,\,\,\,\,\,(p\leq q) \\
              q^2p\,(p^4-q^4/3-\frac{2}{3}p^2q^2)\,\,\,\,\,\,(q\leq p)
            \end{array}
\right.
\label{b9}
\end{equation}
and $q=q_j^{(i)}$ denote the roots of the equation (\ref{b7}).
The above expressions allow a direct computation of the equations
(\ref{b5}), but it will be more instructive to perform the following dimensional 
analysis whose results were numerically verified.
Firstly, from equations (\ref{b5a}) and (\ref{36})
we have  $\tilde{\mu}(0)=\nu=\Omega_{\rm rot}B$ which, 
according to the experimental values of $B$, must be
of order $\Omega_{\rm rot}$ at most. Then from equation (\ref{b5b}) it is not difficult to
realize that $\tilde{\mu}'(0)$ should be  $\sim\Omega_{\rm rot}/w_0$ and
similarly $\tilde{\mu}''(0)\sim \Omega_{\rm rot}/w_0^2$. Thus, the first- and 
second-order corrections in equation (\ref{b4}) turn out to be, respectively,
 of order $\Omega_{\rm rot}/w_0$
and $(\Omega_{\rm rot}/w_0)^2$ which tells us that the Markovian limit corresponds to
$\Omega_{\rm rot}\ll w_0$.

\end{document}